\documentclass[aip, 12pt, nofootinbib]{revtex4}
\usepackage{setspace, subcaption}
\usepackage{amsmath, amssymb, nccmath, physics, tensor}
\usepackage{xparse, graphicx, accents}
\usepackage{mathtools, color, hyperref}
\usepackage[toc,page]{appendix}
\usepackage[mathscr]{euscript}
\newcommand{\lagr}{\mathcal{L}}
\newcommand{\amp}{\mathcal{A}}
\newcommand{\ham}{\mathcal{H}}
\newcommand{\cd}{\mathcal{D}}
\newcommand{\normal}{\mathcal{N}}
\newcommand{\mcal}{\mathcal{M}}
\newcommand{\feyn}{\mathcal{F}}
\newcommand{\realn}{\mathbb{R}}
\newcommand{\ep}{\epsilon}
\newcommand{\kap}{\kappa}

\newcommand{\bx}{\vb{x}}
\newcommand{\bu}{\vb{u}}
\newcommand{\bv}{\vb{v}}
\newcommand{\by}{\vb{y}}

\newcommand{\be}{\begin{equation}}
\newcommand{\ee}{\end{equation}}
\newcommand{\ba}{\begin{eqnarray}}
\newcommand{\ea}{\end{eqnarray}}
\newcommand{\bma}{\left(\begin{array}}
\newcommand{\ema}{\end{array}\right)}

\def\bs{\begin{subequations}}
\def\es{\end{subequations}}

\def\e{\epsilon}

\def\p{\partial}
\newcommand{\Eq}[1]{(\ref{#1})}
\def\lp{\ell_{\rm Pl}}

\begin{document}

\title{An Exact Fermionic Chern-Simons-Kodama State\\ in Quantum Gravity}
\author{Stephon Alexander}
\email{stephon\_alexander@brown.edu}
\author{Tatsuya Daniel}
\email{tatsuya\_daniel@brown.edu}
\affiliation{Brown Theoretical Physics Center, \\ Department of Physics, Brown University, Providence, RI 02912, USA}
\author{Marcell Howard}
\email{mah455@pitt.edu}
\affiliation{Pittsburgh Particle Physics, Astrophysics, and Cosmology Center, \\ Department of Physics and Astronomy, University of Pittsburgh, Pittsburgh, PA 15260, USA}
\author{Morgane K$\ddot{\rm o}$nig}
\email{mkonig@mit.edu}
\affiliation{Department of Physics and Astronomy, Dartmouth College, Hanover, NH 03755, USA}

\begin{abstract}
    The Chern-Simons-Kodama (CSK) state is an exact, non-perturbative wave function in the Ashtekar formulation of classical General Relativity.  In this work, we find a generalized fermionic CSK state by solving the extended gravitational and fermionic Hamiltonian constraints of the Wheeler-DeWitt equation exactly.  We show that this new state reduces to the original Kodama state upon symmetry reduction to FRW coordinates with perturbative fermionic corrections, making contact with the Hartle-Hawking and Vilenkin wave functions of the universe in cosmology.  We also find that when both torsion and fermions are non-vanishing, the wave function possesses a finite amplitude to evade the Big Bang curvature singularity.
\end{abstract}

\maketitle

\section{Introduction}

In various attempts to quantize gravity, the question of background independence (active diffeomorphisms) is a key guiding principle. In non-perturbative quantizations such as loop quantum gravity, causal dynamical triangulations, spin foams and group field theory, only the gravitational degrees of freedom, such as the connection, are elevated to quantization. A criticism of these approaches is that they lack a principle that links the rest of the matter fields in nature to the quantization of gravity.\footnote{The geometrization of matter, which posits that matter fields emerge from conical defects of geometry, has been pursued by Crane, and cosmological applications have been explored \cite{Crane01, Alexander03}.} Supergravity is an approach that was able to link gravity to matter via supersymmetry \cite{VanNieuwenhuizen81, Cheng96, Moniz98, Moniz10, Moniz10-2}. Quantum gravitational approaches, such as string/M-theory, place the graviton and matter fields on the same footing since both arise from excitations from the string vacua.

Another approach in the canonical quantization route is the procedure laid out by Dirac, which quantizes a constrained Hamiltonian system. In the case of general relativity (GR), this invariably leads to the ADM formalism \cite{Arnowitt08}, where the spacetime manifold $\mcal_4$ is decomposed into a family of spacelike 3-dimensional hypersurfaces $\Sigma_t$ which are parametrized by a time coordinate $t$.\footnote{This approach introduces a set of functions made from the time-time and time-space components of the 4-metric $g_{\mu\nu}$: the lapse function $N(\bx,t)$ and the shift function $N^a(\bx,t)$. The lapse function relates variations in coordinate time $t$ to those of the proper time as one follows a curve that lies normal to $\Sigma_t$. The shift vector, on the other hand, describes the variations in a spatial point as one moves along a curve that lies tangential to $\Sigma_t$.} 

This approach is also not without its problems. The Hamiltonian and momentum conjugate enter the action as constrained quantities.  The resulting quantization of the Hamiltonian constraint gives the Wheeler-DeWitt equation, which is a second-order hyperbolic functional differential equation, corresponding to an infinite number of degrees of freedom at each spatial point. This makes any calculation cumbersome at best and ill-defined at worse. The Ashtekar formulation of GR \cite{Ashtekar86} provides one way to tame some of the issues pertaining to this naive canonical quantization picture.

The Ashtekar formalism (for which we will delve into more detail in the next section) recasts the dynamical variables in GR from the metric to a Yang-Mills gauge field over the $SU(2)$ gauge group. This reduces the Hamiltonian to a much simpler and far more manageable form where one is dealing with a polynomial in the gauge field and its canonical partner. This allows the Wheeler-DeWitt equation to be solved exactly, amounting to finding the ground state of the quantized theory; with the inclusion of a cosmological constant, the solution is known as the Chern-Simons-Kodama (CSK) state. The fact that the CSK state is an exact, non-perturbative solution makes it a promising candidate for describing quantum gravity with a cosmological constant. More broadly, the Ashtekar variables also provide a natural way to incorporate couplings to matter fields. This leads to a potentially auspicious avenue of working within a universal, non-perturbative framework for treating both the geometry and matter sectors on equal footing, as the general theory of relativity would have us do.

Since all of the known matter in the Standard Model are fermions, we seek to find a new CSK state that includes fermionic matter on the same footing as gravity (for inclusions of bosonic matter see \cite{Alexander04}). In this work, we explore a quantization of gravity with the inclusion of fermionic matter by solving both the gravitational and fermionic Hamiltonian constraint. We find an exact wave function that has interesting connections to the CSK state with the inclusion of torsion.  We then seek to make contact with the Hartle-Hawking/Vilenkin wave functions of quantum cosmology from this exact wave function.  

This paper is organized as follows: in Section \ref{sec:2}, we provide a brief overview of the Ashtekar formalism and derive the Kodama state from the quantum Hamiltonian constraint. In Section \ref{sec:3}, we add fermions and find we can still solve the modified Hamiltonian constraint exactly. In Section \ref{sec:4}, we add torsion into the picture and find that the original Hartle-Hawking (or Vilenkin) state is the leading term to the resulting wave function when we expand around the fermion fields. We conclude with some remarks and directions for future work.


\section{The Ashtekar Formalism and Kodama State\label{sec:2}}
In pursuit of a Wheeler-DeWitt quantization of gravity, it is instructive to understand how the Ashtekar connection and the resulting  Hamiltonian, diffeomorphism, and gauge constraints emerge from a manifestly covariant 4D theory of gravity.  In the Ashtekar formalism \cite{Ashtekar86}, gravitational dynamics on a four-dimensional manifold ${\cal M}_4$ is not described by a metric $g_{\mu\nu}$ but\footnote{We use the mostly plus metric signature, i.e. $\eta_{\mu\nu} = (-,+,+,+)$ in units of $c = 1$. We use boldface letters $\bx$ to indicate 3-vectors, and we use $x$ to denote 4-vectors. Conventions for curvature tensors, covariant and Lie derivatives are all taken from Carroll \cite{Carroll04}. Greek indices ($\mu,\nu,\ldots)$ denote spacetime indices, Latin indices $(a, b,\ldots)$ denote spatial indices, and Latin indices $(i, j, \ldots)$ denote indices for the internal space.}, rather, a real-valued gravitational field $e_\mu^i(x)$, mapping a vector $v^\mu$ in the tangent space of ${\cal M}_4$ at the point $x$ into Minkowski spacetime $M_4$ (with metric $\eta_{ij} = {\rm diag}(-1,1,1,1)_{ij}$). Locally, the metric on ${\cal M}_4$ is $g_{\mu\nu} = \eta_{ij}e_\mu^ie_\nu^j$.

The Lorentz connection $\tensor{\omega}{_{\mu i}^j}$ is $\tensor{\omega}{_i^j} \equiv \tensor{\omega}{_{\mu i}^j} \dd{x^\mu}$, $\dd{\tensor{\omega}{_i^j}}\equiv \p_\mu \tensor{\omega}{_{\nu i}^j}\dd{x^\mu} \wedge \dd{x^\nu}$ is the exterior derivative, and the curvature of $\omega$ is $\tensor{R}{_i^{j}} = \dd{\tensor{\omega}{_i^{j}}} + \tensor{\omega}{_i^{k}} \wedge \tensor{\omega}{_k^{j}}$. The action of self-dual gravity is (up to the gravitational constant $G$)
\be\label{act}
S = \frac{1}{32\pi G}\int_{{\cal M}_4} \qty[*(e^i\wedge e^j)\wedge R_{ij}  + ie^i\wedge e^j\wedge R_{ij} - \frac{\Lambda}{6}\ep_{ijk\ell}e^i\wedge e^j\wedge e^k\wedge e^\ell]\,,
\ee
where $*$ is the Hodge dual, the first term is the Hilbert--Palatini action and the second is the Holst term (proportional to the first Bianchi identities in the absence of torsion).

Here we are interested in the Hamiltonian formulation in Ashtekar variables \cite{Ashtekar86, Ashtekar87}. In the gauge choice $e^0_\mu = 0$, it is convenient to define the densitized triad $E^{a}_{i} = \ep_{ijk}\ep^{abc} e^j_be^k_c$, which is conjugate to the self-dual connection
\be\label{sdc}
A_a^i(x) \equiv -\tfrac12 \tensor{\ep}{^{ij}_{k}}\tensor{\omega}{_{aj}^{k}} - i\tensor{\omega}{_{a0}^{i}}.
\ee
As the Lorentz connection (and, in particular, the spin connection $\Gamma_a^i \equiv -\tfrac12 \tensor{\ep}{^{ij}_{k}}\tensor{\omega}{_{aj}^{k}}$) is real, $A$ is complex-valued and obeys the reality conditions (for a discussion, see e.g. \cite{Kuchar93})
\be\label{rc}
A_a^i +  A_a^{i*}= 2\Gamma_a^i[E]\,,
\ee
where $*$ denotes complex conjugation and the spin connection solves the equation $\dd{e} + \Gamma[E]\wedge e = 0$.

The Poisson bracket of the elementary variables $A$ and $E$ is
\be\label{comm}
\pb{A_a^i(\bx)}{E_j^b(\by)} = i8\pi G\delta_a^b\delta_j^i\delta({\bx} - {\by})\,.
\ee
Introducing the "magnetic" field and the gauge field strength
\ba\
B^{ai}&\equiv& \frac{1}{2}\e^{abc}F_{bc}^i\,,\\
F_{ab}^k &=& \p_a A_b^k - \p_b A_a^k + (8\pi G)\tensor{\e}{_{ij}^{k}}A_a^iA_b^j\,\label{F},
\ea
one can show that the Hamiltonian scalar constraint following from Eq.~\Eq{act} is
\ba\label{sca}
\ham &\equiv& \ep_{ijk} E^{ai}E^{bj} \qty(F_{ab}^k + \frac{\Lambda}{3}\e_{abc}E^{ck}) = 0,
\ea
where $\Lambda$ is a cosmological constant (of any sign) and $\times$ is the vector spatial product defined as $({\bu}\times {\bv})^a = \tensor{\ep}{^a_{bc}}u^bv^c$. In exterior algebra notation, the gauge field is $A \equiv A_a \dd{x^a} \equiv A_a^i \tau_i \dd{x^a}$ ($\tau_i$ being an $su(2)$ generator), the covariant derivative is $\cd \equiv \dd + (8\pi G)A\ \wedge$, and Eq.~\Eq{F} can be compactly recast as $F = \dd{A} + (8\pi G)A\wedge A$. Under a local gauge transformation, the Ashtekar connection transforms as
\be\label{gt}
A \rightarrow A'= gAg^{-1} - g^{-1}\dd{g}\,,
\ee
where $g({\bx})$ is an element of the gauge group of gravity $\mathcal{G} = SU(2)$. Let $\mathcal{G}_0$ be the subgroup of \emph{small gauge transformations}, i.e., local transformations continuously connected to the identity. Its elements are of the form $g_0 = \exp[-i\theta^i(\bx)\tau_i]$, where $\theta^i(\bx)$ are some functions on a spatial slice of ${\cal M}_4$. Pure gauge configurations $g^{-1}\dd{g}$ are equivalent to the flat gauge $A = 0$.

The full invariance group of the theory is the semidirect product of the diffeomorphism and gauge groups. Invariance under small gauge transformations is guaranteed by the Gauss constraint
\be\label{gau}
{\cal G}_i \equiv \cd_a E^a_i = \p_a E^a_i + (8\pi G)\e_{ijk}A_a^j E^{ak}=0,
\ee
while spatial diffeomorphism invariance is imposed by the vector constraint
\be\label{vec}
{\cal V}_a \equiv (E_i\times B^i)_a = E_i^b F_{ab}^i = 0\,.
\ee
The total Hamiltonian is a linear combination of the constraints; up to constants, $H = (8\pi G)^{-1}\int_{{\cal M}_3} \dd[3]{x} (N\ham + N^a{\cal V}_a + \lambda^j{\cal G}_j)$, where ${\cal M}_3$ is the spatial submanifold and $N$, $\lambda^j$, and $N^a$ are Lagrange multipliers (in particular, $\lambda^j$ is a generator of $su(2)$).

Now, let us construct the CSK state by following the example of \cite{Kodama90} to solve the Wheeler-DeWitt equation. We have the Hamiltonian
\begin{equation}
    \ham_{WDW} = \ep_{ijk}E^{ai}E^{bj}\qty(F^k_{ab} + \frac{\Lambda}{3}\ep_{abc}E^{ck}), \label{eq:ham} 
\end{equation}
which acts on some wave function $\psi[A]$, and we want to find the form of $\psi[A]$ that is annihilated by (\ref{eq:ham}). Applying the regular canonical quantization procedure, i.e.
\begin{equation}
    \hat{E}^{ai}\rightarrow -8\pi G\hbar\fdv{A_{ai}},
\end{equation}
the annihilation of the quantum state becomes
\begin{equation}
    \widehat{\ham}_{WDW}\psi[A] = (8\pi G\hbar)^2 \ep_{ijk}\fdv{A_{ai}}\fdv{A_{bj}}\qty(F^k_{ab} - \frac{8\pi G\hbar\Lambda}{3}\ep_{abc}\fdv{A_{ck}})\psi[A] = 0.
\end{equation}
If we assume that the field strength is self-dual, then $F^k_{ab} = -\frac{\Lambda}{3}\ep_{abc}E^{ck}$ so
\begin{equation}
    \ep_{abc}\fdv{\psi}{A_{ck}} = \frac{3}{8\pi G\hbar\Lambda}F^k_{ab}\psi[A].
\end{equation}
Contracting both sides with $\ep^{dab}$ gives us
\begin{equation}
    2\delta^d_c\fdv{\psi}{A_{ck}} = \frac{3}{\ell_{\rm Pl}^2\Lambda}\ep^{dab}F^k_{ab}\psi[A]\Leftrightarrow \fdv{\psi}{A_{ai}} = \frac{3}{2\ell^2_{\rm Pl}\Lambda}\ep^{abc}F^i_{bc}\psi[A],
\end{equation}
where $\ell_{\rm Pl}^2 = 8\pi G\hbar$ is the Planck length. Recognizing the term multiplying the wave function to be the Chern-Simons functional, we can write down the exact solution to the Wheeler-DeWitt equation as being
\begin{equation}
    \psi_K[A] \equiv \mathcal{N}\exp(-\frac{3}{2\ell_{\rm Pl}^2\Lambda}\int Y_{CS}[A]),
\end{equation} 
where $\mathcal{N}$ is some normalization constant independent of the gauge field and 
\begin{equation}
    Y_{CS}[A] = \Tr[A\wedge\dd{A} + \frac{2}{3}A\wedge A\wedge A] = -\frac{1}{2}\qty(A^i\dd{A^i} + \frac{1}{3}\ep_{ijk}A^iA^jA^k)
\end{equation}
is the Chern-Simons functional, with the trace taken in the Lie algebra. It can be said that the WKB semiclassical limit of the CSK state is de Sitter spacetime\footnote{See \cite{Witten03} for criticisms of this view.} \cite{Smolin02}, with
\be
A_a^{i} = i\sqrt{\frac{\Lambda}3}\,e^{\sqrt{\frac{\Lambda}{3}}\, t}\delta_a^i\,,\qquad E_i^a = e^{2\sqrt{\frac{\Lambda}{3}}\, t}\delta^a_i\,.
\ee

Now that we have the CSK state solely in terms of the gravitational connection and the cosmological constant, we would like to explore a full non-perturbative state that also includes the fermionic Hamiltonian.  

\section{The Fermionic CSK Solution\label{sec:3}}

To find the Hamiltonian constraint associated with fermions covariantly coupled to gravity, we start with the covariant fermionic Lagrangian \cite{Jacobson88, Morales94}: 

\begin{equation}
	\lagr = ee^{aAC}\tensor{e}{^{bB}_{C}}\feyn_{abAB} - 2e\Lambda + \sqrt{2}ee^{aAB}\bar{\xi}_BD_a\xi_A,
\end{equation}
where $e^{aAB}$ is the spinorial representation of the tetrads, $\amp_{aAB}$ is the 4D gauge field with curvature tensor $\feyn_{abAB}$, $\xi$ is a 2-component spinor, and the covariant derivative acting on spinors is
\begin{equation}
    D_a\xi_A \equiv \p_a\xi_A + (8\pi G)\tensor{\amp}{_{aA}^B}\xi_B.
\end{equation}
Next, we employ the ADM variables by first decomposing the full 4D manifold $\mcal_4 = \realn\times\Sigma$. We define a scalar function $t$ (that acts as our time coordinate) and vector field $t^a$ such that
\begin{equation}
	t^a\nabla_a t = 1,
\end{equation}
where $\nabla$ is the torsion free $e^{aAB}$-compatible connection. We also pick $t$ under the constraint that hypersurfaces of constant $\Sigma_t$ are spacelike and $t^a$ is timelike. We define a new timelike unit vector $n^a$ such that $n^an_a = -1$ at each point on $\Sigma_t$. We can define a unit spinor by
\begin{equation}
	n^{AB} \equiv n_ae^{aAB}.
\end{equation}
The induced metric on $\Sigma_t$ is then given by
\begin{equation}
	\gamma_{ab} \equiv g_{ab} + n_an_b,
\end{equation}
and so every tensor field can be decomposed into a part that is orthogonal to $\Sigma_t$ and a component that is parallel to it. In particular, we have
\begin{equation}
	t^a = Nn^a + N^a,
\end{equation}
where $N$, $N^a$ are the previously mentioned lapse and shift vector respectively, and $n^aN_a = 0$. Now, the object $-i\sqrt{2}n^{AB}$ defines a Hermitian metric with the property
\begin{equation}
	n^{AB}n_{CB} = \frac{1}{2}\delta^A_C,
\end{equation}
which has the action of mapping spinors to their Hermitian conjugate, i.e. $\xi_A^\dagger \equiv -i\sqrt{2}\tensor{n}{_A^B}\bar\xi_B$. Next, we can extract the spatial part of $e^{aAB}$ by defining
\begin{equation}
	E^{aAB} \equiv i\sqrt{2}e^{aC(A}\tensor{n}{^{B)}_C},
\end{equation}
and we then have the identity
\begin{equation}
	e^{aAB} = n^an^{AB} + i\sqrt{2}E^{aAC}\tensor{n}{^B_C}.
\end{equation}
From these definitions, we get $n_aE^{aAB} = 0\ $and $\tensor{E}{^a_A^A} = 0$, where we raise and lower the spinorial indices using the 2D Levi-Civita symbol $\ep_{AB}$, which acts as a metric on the vector space of spinors. Now we need only to plug these definitions into the action to get
\begin{equation}
	S_{EH} = \frac{1}{2\kap}\int\dd{t}\int_{\Sigma_t}\dd[3]{x}NE\qty[i\sqrt{2}n^aE^{bAB}F_{abAB} - E^{aAC}\tensor{E}{^{bB}_C}F_{abAB} - 2\Lambda],
\end{equation}
where we have introduced the notation $\kap \equiv 8\pi G$, $E^{aAB}$ projects out the timelike component of $\feyn_{abAB}$, leaving only the spatial components $F_{abAB}$. Next, we write $n^a = (t^a - N^a)/N$, and replug it into the action to find
\begin{ceqn}
	\begin{align}
		S_{EH} = \frac{1}{2\kap}\int\dd[4]{x}\qty[i\sqrt{2}\Tr[\tilde{E}^bt^aF_{ab}] - i\sqrt{2}N^a\Tr[\tilde{E}^bF_{ab}] - N\Tr[\tilde{E}^aE^bF_{ab}] + 2NE\Lambda],
	\end{align}
\end{ceqn}
where overhead tildes denote densitized quantities, the trace is taken in the $su(2)$ and the $t^aF_{abAB}$ term is found to be
\begin{ceqn}
	\begin{align}
		t^aF_{abAB} = \lagr_t A_{bAB} - \cd_b A_{tAB},
	\end{align}
\end{ceqn}
where $A_{tAB} \equiv t^aA_{aAB}$, $\lagr_t$ is the Lie derivative along $t^a$ (which we will henceforth denote as overhead dots), and $\cd_b$ is the covariant derivative with connection $A_{aAB}$. We are then left with
\begin{equation}
	S_{EH} = \frac{1}{2\kap}\int\dd{t}\int_{\Sigma_t}\dd[3]{x}\qty[i\sqrt{2}(\Tr\dot{A}_a\tilde{E}^a + \Tr A_t\cd_a\tilde{E}^{a} - N^a\Tr\tilde{E}^bF_{ab}) - N(\Tr\tilde{E}^aE^bF_{ab} - 2E\Lambda)],
\end{equation}
where we integrated by parts on the covariant derivative term. For a discussion on the neglected boundary terms, see 4.4 in \cite{Jacobson88-2}. Next, we focus on the matter sector for which (after some tedious algebra) the action is
\begin{equation}
	S_f = \int\dd[4]{x}\qty[\widetilde{\Pi}^A\dot{\xi}_A + \tensor{A}{_{tA}^B}\widetilde{\Pi}^{A}\xi_{B} - N^a\widetilde{\Pi}^A\cd_a\xi_A + i\sqrt{2}N\tilde{E}^{aAB}\Pi_B\cd_a\xi_A].
\end{equation}
This brings the total action to the form
\begin{equation}\label{tot_act}
	\begin{split}
	S &= \int\dd{t}\int_{\Sigma_t}\dd[3]{x}\left[\frac{i\sqrt{2}}{2\kap}\Tr(\dot{A}_a\tilde{E}^a) + \widetilde{\Pi}^A\dot{\xi}_A + A_{tAB}\qty(\frac{i\sqrt{2}}{2\kap}\cd_a\tilde{E}^{aAB} + \xi^{(B}\widetilde{\Pi}^{A)}) \right. \\ &\left. + \underaccent{\tilde}{N}\qty(-\frac{1}{2\kap}\qty(\Tr(\tilde{E}^a\tilde{E}^bF_{ab}) + 2E\Lambda) + i\sqrt{2}\tilde{E}^{aAB}\widetilde{\Pi}_B\cd_a\xi_A) + N^a\qty(-\frac{i\sqrt{2}}{2\kap}\Tr(\tilde{E}^{b}F_{ab}) + \widetilde{\Pi}^A\cd_a\xi_A)\right],
	\end{split}
\end{equation}
where $\underaccent{\tilde}{N} \equiv N/E$. We shall henceforth drop all overhead tildes with the understanding that all quantities are densitized. Finally, the sympletic structure for the spinor field and its conjugate partner is 
\begin{equation}
    \pb{\xi^A(\bx)}{\Pi_B(\by)} = \delta^A_B\delta^3(\bx - \by).
\end{equation}
The introduction of the fermionic fields has resulted in modifications to the scalar, vector, and Gauss constraints that we saw earlier. Given the fact that $N$ (or in this case $N/E$), $N^a$, and $A_{tAB}$ all enter into the action as Lagrange multipliers, we are again inspired to find quantum states that can be simultaneously annihilated by all three new constraints. First we start off with the scalar constraint but with the addition of a new Hamiltonian
\begin{equation}
    \ham_f = i\sqrt{2}(\cd_a\xi)_AE^{aAB}\Pi_B,
\end{equation}
with
\begin{equation}
    E^{aAB}(x) = -\frac{i}{\sqrt{2}}E^{ai}(x)\sigma_i^{AB}, \hspace{0.5cm}\tensor{A}{_{aA}^B}(x) = A_a^i(x)\tensor{\tau}{_{iA}^B},
\end{equation}
where $\sigma_i$ are the Pauli matrices, and $\tau_j = -i\sigma_j/2$ are the $su(2)$ generators. We are interested in finding a state vector $\Psi[A,\xi]$ such that
\begin{equation}
    \qty(\widehat{\ham}_{WDW} + \widehat{\ham}_f)\Psi[A,\xi] = 0,
\end{equation}
i.e. it is simultaneously annihilated by the Wheeler-DeWitt and fermionic Hamiltonian constraints. Next we write
\begin{equation}
    \qty[\frac{1}{2\kap}\ep_{ijk}\hat{E}^{ai}\hat{E}^{bj}\qty(\hat{F}^k_{ab} + \frac{\Lambda}{3}\ep_{abc}\hat{E}^{ck}) + 2\widehat{(\cd_a\xi)}_A\hat{E}^{ai}\sigma_i^{AB}\widehat{\Pi}_B]\Psi[A,\xi] = 0,
\end{equation}
where we used $E = \frac{1}{3!}\ep_{ijk}\ep_{abc}E^{ai}E^{bj}E^{ck}$ and we have suppressed the Dirac indices. Assuming non-degeneracy of the triads, we can factor out a triad, which leads to a Hamiltonian that is a little simpler:
\begin{equation}
    \qty[\frac{1}{2\kap}\ep_{ijk}\hat{E}^{bj}\qty(\hat{F}^k_{ab} + \frac{\Lambda}{3}\ep_{abc}\hat{E}^{ck}) + 2\widehat{(\cd_a\xi)}_A\sigma_i^{AB}\widehat{\Pi}_B]\Psi[A,\xi] = 0.
\end{equation}
We then apply the usual canonical quantization scheme
\begin{equation}
    \hat{E}^{ai}\rightarrow -\ell_{\rm Pl}^2\fdv{A_{ai}}, \hspace{0.5cm}\widehat{\Pi}_A\rightarrow - i\hbar\fdv{\xi^A}.
\end{equation}
Taking our ansatz for the wave function to be $\Psi[A,\xi] = \Omega[A]\psi_K[A]e^{\alpha_A\xi^A}$, as well as neglecting higher-order derivative terms because the wave function is occupying its ground state, reduces us to the following constraint
\begin{equation}
    \frac{1}{2}\ep_{ijk}F^j_{ab}\fdv{\Omega}{A_{bk}} + 2i(\cd_a\xi)_A\sigma_i^{AB}\alpha_B\Omega[A] = 0,
\end{equation}
where $\alpha_A$ is a constant spinor. Next, we employ symmetry reduction arguments, i.e. we restrict our solution space to only include gauge fields which are homogeneous and isotropic, i.e.
\begin{equation}\label{mss}
    A_{ai} = iA(t)\delta_{ai}\Rightarrow F^j_{ab} = -\kap A^2\tensor{\ep}{^j_{ab}}.
\end{equation}
This brings our constraint equation to the form
\begin{equation}\label{mss2}
    A^2\frac{1}{\Omega[A]}\fdv{\Omega}{A}\delta_{ai} = \frac{1}{4\pi G}(\cd_a\xi)_A\sigma_i^{AB}\alpha_B,
\end{equation}
where we used the fact that
\begin{equation}
    \fdv{\Omega}{A_{ai}} = -i\delta^{ai}\fdv{\Omega}{A}.
\end{equation}
From here, we can solve the constraint equation exactly by simply integrating our result. Doing so after taking the trace on both sides, we get
\begin{equation}
    \Omega[A] = \Omega_0\exp(\frac{1}{12\pi G} \int^A\frac{(\cd_a\xi)_{A'}\sigma_a^{A'B}\alpha_B}{A'^2}\dd{A'}),
\end{equation}
where the $\xi$ in the exponential is said to be fixed, and we use the notational index $A'$ to communicate that we are using the $A'$ gauge field. The exact wave function solution for the combined gravity and fermion system is\footnote{We can re-plug this term into the (symmetry-reduced) constraint in order to see if our original justification to ignore the third derivative term was valid. Doing so yields
\begin{equation*}
    \fdv[3]{\Omega}{A} \simeq \order{\frac{\hbar^3}{A^3}},
\end{equation*}
meaning for a fixed value for $\xi$, in the limit of small $\hbar$ and large $A$, this term is indeed negligible.}
\begin{equation}
    \Psi[A,\xi] \equiv \Omega[A]\psi_K[A]e^{\alpha_A\xi^A} = \Omega_0\exp(\frac{1}{4\pi G} \int^A\frac{(\cd_a\xi)_{A'}\sigma_a^{A'B}\alpha_B}{3A'^2}\dd{A'})\psi_K[A]e^{\alpha_A\xi^A}. \label{eq:wftot}
\end{equation}

We see that our wave function (\ref{eq:wftot}) has the structure of a product wave function of the pure CSK state and an integral part $\Omega[A]$, which is suppressed for large values of the gravitational connection. In order to make connection with cosmology, we will pursue studying this exact state in a mini-superspace (FRW) background, assuming non-vanishing torsion.

\section{Mini-superspace reduction of the Fermionic State with Torsion}\label{sec:4}

\begin{figure}
    \centering
    \begin{subfigure}[l]{0.49\linewidth}
        \centering
        \includegraphics[width=\linewidth]{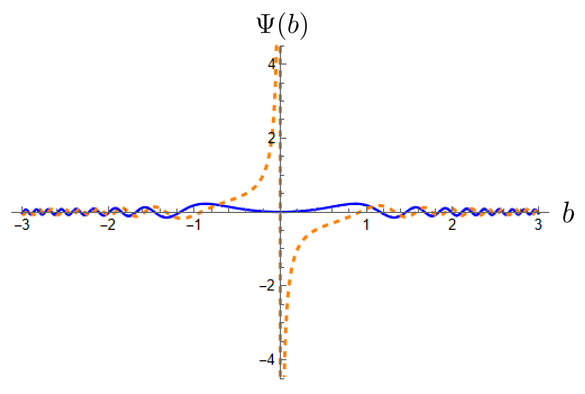}
    \end{subfigure}
    \begin{subfigure}[l]{0.49\linewidth}
        \centering
        \includegraphics[width=\linewidth]{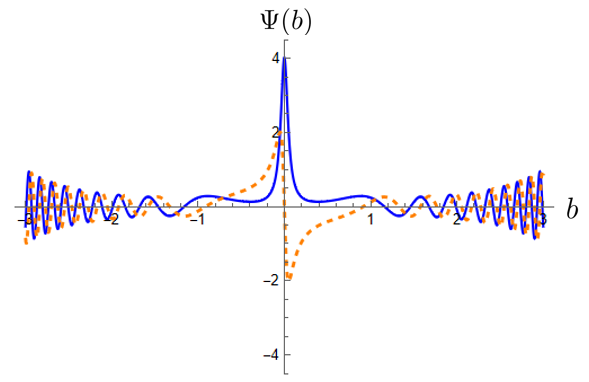}
    \end{subfigure}
    \caption{The real (solid blue) and imaginary (dashed orange) parts of the full wave function (\ref{eq:wffull}) for $\alpha_A\xi^A = 1$.  The plot on the left is for $c = 0$, while on the right $c = 0.05$.}\label{fig:1}
\end{figure}

\begin{figure}
    \centering
    \begin{subfigure}[l]{0.49\linewidth}
        \centering
        \includegraphics[width=\linewidth]{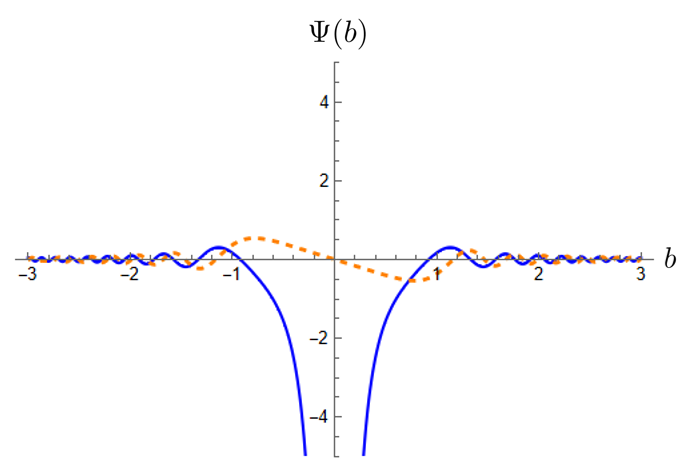}
        \caption{}
    \end{subfigure}
    \begin{subfigure}[l]{0.49\linewidth}
        \centering
        \includegraphics[width=\linewidth]{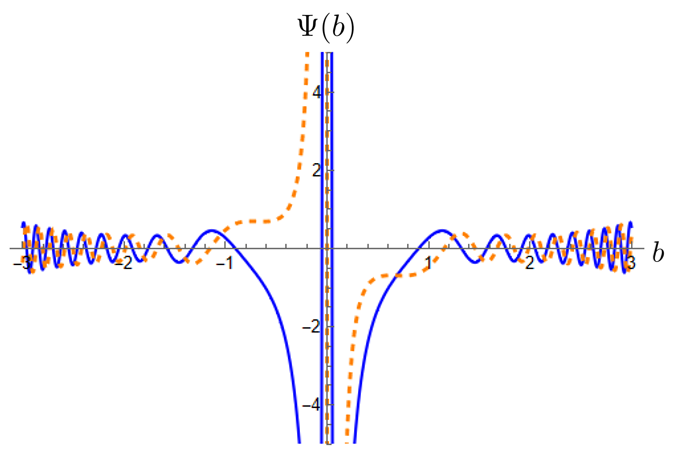}
        \caption{}
    \end{subfigure}
    \caption{The real (solid blue) and imaginary (dashed orange) parts of the full wave function (\ref{eq:wffull}) for $\alpha_A\xi^A = 2$. The plot on the left is for $c = 0$, while on the right $c = 0.05$. For even values of the spinor product $\alpha_A\xi^A$ and when $c = 0$, the real part of the wave function diverges as $-b \rightarrow 0$ while the imaginary part remains finite; this phenomenon is switched for odd values (see Fig.~\ref{fig:1}).}\label{fig:2}
\end{figure}

 It is well known \cite{Jacobson88} that in the first-order and Ashtekar formalism, fermions source torsion. We are interested in exploring the relationship between fermions and a symmetry-reduced torsion at the level of our full wave function, similar to what was considered in \cite{Magueijo19} (see \cite{Bojowald08, Dolan10, Myrzakulov19} for other treatments of torsion and fermions). We relax the torsion-free condition on the spin connection while keeping the metric compatibility constraint (thus the spin connection is still anti-symmetric in its internal indices). Recall the definition of the torsion
\begin{equation}
    T^i \equiv \dd{e^i} + \tensor{\omega}{^i_j}\wedge e^j,
\end{equation}
where we have suppressed the spacetime index for the time being. On a homogeneous and isotropic spacetime, the vierbein, torsion, and spin connection become
\begin{equation}
    e^0 = \dd{t},\hspace{0.5cm} e^i = a\dd{x^i},
\end{equation}
\begin{equation}\label{torsion}
    T^0 \equiv 0, \hspace{0.5cm} T^i \equiv -T(t)e^0e^i + P(t)\tensor{\ep}{^i_{jk}}e^je^k,
\end{equation}
\begin{equation}
    \tensor{\omega}{^i_0} = \qty(\frac{\dot{a}}{a} + T)e^i \equiv g(t)e^i,\hspace{0.5cm}\omega^{ij} = -P\ep^{ijk}e^k,
\end{equation}
where we have defined $g(t) \equiv \qty(\frac{\dot{a}}{a} + T)$. We note here that $g(t)$ is a real-valued function that plays the role of the Hubble parameter. Upon plugging the spin connection to (\ref{sdc}), we get the particularly simple form for the gauge field
\begin{equation}
	A^i_a = a\qty(P - ig)\delta^i_a \equiv (ib + c)\delta^i_a,
\end{equation}
where $-b = \dot{a} + aT$ and $c = aP$. Now we can just plug this into (\ref{mss2}): 
\begin{equation}
	\frac{1}{\Omega(b)}\dv{\Omega}{b} = -\frac{1}{12\pi G}\frac{\cd_a\xi_A\sigma^{AB}_a\alpha_B}{(ib + c)^2}.
\end{equation}
Now we take the spinor field to be homogeneous\footnote{See \cite{Ochs94, Armend03, Vakili05} for further exploration of spinor fields on a cosmological background.}, $\xi = \xi(t)$, which enables us to solve for the exact form of the wave function,
\begin{equation}
	\Omega(b) = \Omega_0(ib + c)^{-\alpha_A\xi^A},
\end{equation}
where we used $\xi^A \equiv \ep^{AB}\xi_B = -\xi_A$ and $\Omega_0$ is a normalization constant. The full wave function is then
\begin{equation} 
	\Psi_b(b,c,\xi) = \Omega_0(ib + c)^{-\alpha_A\xi^A}\psi_K(b)e^{\alpha_A\xi^A}, \label{eq:wffull}
\end{equation}
where the symmetry-reduced Kodama State becomes
\begin{equation}
    \psi_K(b) = \frac{1}{\sqrt{2\pi}}\exp(-\frac{3V_c}{\Lambda\lp^2}(ib + c)^3),
\end{equation}
where we have chosen the normalization constant $\sqrt{2\pi} \equiv \normal^{-1}$ because we require the wave function to be delta-function normalizable, and $V_c$ is the volume of the 3D hypersurface. The fact that we have obtained an exact solution for the full wave function in the form of (\ref{eq:wffull}) is intriguing, and it is plotted in Figs.~\ref{fig:1} and \ref{fig:2}.

 Recently, Magueijo discovered \cite{Magueijo20} that with different choices of contours, the Kodama State is the Fourier dual to the Hartle-Hawking wave function \cite{Hartle83} and Vilenkin (or tunneling) states \cite{Vilenkin89}. In \cite{Alexander21}, this analysis was extended to include both torsion as well as beyond mini-superspace solutions to the Wheeler-DeWitt equation. We would like to see if our full fermionic wave function can make contact with these previous results in order to better interpret the solution we have. Reducing our full wave function (\ref{eq:wffull}) to mini-superspace, we are able to read off the commutation relations between the connection and its electric field:
\begin{align}
    \comm{\hat{b}}{\hat{a}^2} = \frac{i\lp^2}{3V_c}.
\end{align}

In the numerical plots of the wave function (\ref{eq:wffull}) above, we see an interesting generic behavior due to the presence of fermions and torsion. First, we find that, in general, the wave function has divergences as we approach vanishing scale factor, signaling that this form of the wave function is ill-defined at the classical Big Bang singularity. These divergences occur when torsion is vanishing and the fermion amplitude is non-vanishing.  However, this divergence in the wave function is dynamically regulated when we have both non-vanishing torsion and fermion amplitudes.  This is suggestive of a quantum version of the classical statement that fermion currents source torsion \cite{Jacobson88, Albertini22}. The fact that the wave function has non-vanishing probability at the classical Big Bang suggests that it can realize a quantum bounce, which was explored by Geilen and Magueijo \cite{Gielen22}. We now study the wave function analytically.

Because the action is manifestly real-valued, we can replace $\int Y_{CS}\rightarrow i\Im\int Y_{CS}$, which brings the Kodama state to the form
\begin{equation}
    \psi_K(b) = \normal\exp(\frac{3iV_c}{\Lambda\ell^2_{\rm Pl}}(b^3 - 3bc^2)).
\end{equation}
Moving into the $a^2$ representation, and taking our contour to be the real number line, the Fourier transform of our fermionic wave function (\ref{eq:wffull}) becomes:
\begin{align}
    \Psi_{a^2}(a^2, c, \xi) &= \frac{\sqrt{3V_c}}{\ell_{\rm Pl}}\int_{\realn} \dfrac{\dd{b}}{\sqrt{2\pi}}\exp\qty(-\frac{3iV_c}{\ell^2_{\rm Pl}}a^2b)\Psi_b(b, c, \xi) \\
    &= \frac{\sqrt{3V_c}}{\ell_{\rm Pl}}e^{\alpha\cdot\xi} \int_{\realn}\dfrac{\dd{b}}{\sqrt{2\pi}}e^{-\frac{3iV_c}{\ell^2_{\rm Pl}}a^2b}\qty[1 - \alpha_A\xi^A\ln(ib + c) + \ldots]\psi_K(b) \\ &= \normal'{\rm Ai}(-z)e^{\alpha_A\xi^A} - \normal''[\xi]\int_{\realn}\dd{b}e^{-\frac{3iV_c}{\Lambda\ell_{\rm Pl}^2}(b^3 - 3bc^2) - \frac{3iV_c}{\ell_{\rm Pl}^2}a^2b}\ln(ib+c) + \ldots, \label{eq:62}
\end{align}
where we expanded $\Psi_b(b,c,\xi)$ about the spinor product $\alpha_A\xi^A$ using the fact that $x^{-p} = \sum_{n=0}^{\infty} \frac{p^n(-\ln(x))^n}{n!}$, with
\begin{align}
    z &= \qty(\frac{9V_c}{\Lambda\ell^2_{\rm Pl}})^{2/3}\qty(c^2 + \frac{\Lambda a^2}{3}), \\
    \normal' &= \normal\frac{\sqrt{6\pi V_c}}{\ell_{\rm Pl}}\qty(\frac{\Lambda\ell^2_{\rm Pl}}{9V_c})^{1/3}, \\
    \normal''[\xi] &= \frac{\normal}{\lp}\sqrt{\frac{3V_c}{2\pi}}e^{\alpha\cdot\xi}\alpha_A\xi^A.
\end{align}

Evaluation of the corrections to the zeroth order term in (\ref{eq:62}) shows that all higher terms are much smaller than the previous one if $\alpha_A\xi^A \ll 1$. We find this also holds when we instead select our contour of integration to be the negative imaginary number line and the positive real number line; in this case, the zeroth order term of (\ref{eq:62}) is the Vilenkin wave function instead of the Hartle-Hawking wave function \cite{Magueijo20}. The convergence of (\ref{eq:62}) using this complex contour is in line with what is argued in \cite{Halliwell89, Halliwell89-2, Halliwell90}, where the functional integral over the matter fields describing the wave function of the universe is closely related to describing quantum field theory in curved spacetime.

Equation (\ref{eq:62}) shows that the inclusion of fermions can be interpreted as corrections to the Hartle-Hawking (or Vilenkin) wave function, at least when restricting to homogeneous and isotropic metrics, provided the scalar product $\alpha_A\xi^A$ is sufficiently small.

\section{Discussion\label{sec:5}}

We have found an exact solution to the quantized Wheeler-DeWitt equation when one introduces matter fields by working in the Ashtekar formalism. This approach replaces the traditional second-order hyperbolic functional differential equation with a cubic polynomial equation. Our wave function can be generalized to include all fermionic species of the Standard Model by replacing $\bar{\xi}^B\cd_a\xi^A\rightarrow \sum_f\bar{\xi}^B_f\cd_a\xi^A_f$ in the action, where $f$ labels the fermionic species.

A striking feature of this new wave function (\ref{eq:wftot}) is that it cannot be written solely as a product of gravity and fermion wave functions, even though the Hamiltonian is a sum of the gravity and fermion sectors. Instead, the wave function requires an integration of the fermionic configuration convolved with the connection. It will be interesting to numerically simulate this state in the presence of a propagating fermion field.  In order to make progress and make contact with cosmology, we find that the mini-superspace approximation of our fermionic wave function gives back the Hartle-Hawking and Vilenkin wave functions of the universe of quantum cosmology, with perturbative corrections that depend on the spinor and torsion.  

The cosmological realization of our wave function provides new solutions that has no divergences at what would be a classical Big Bang curvature singularity\footnote{Provided that the parity-even or -odd component of the torsion has a dependence on the scale factor that is $\propto a^{-p}$ for $p\geq 1$.}, suggesting a quantum gravitational resolution to that singularity. It has been expected for some time that fermions sourced by torsion can semi-classically resolve the Big Bang singularity, and we plan to explore how our wave function might be related to these results \cite{Alexander09, Poplawski12}. These new cosmological solutions are reminiscent of quantizing fermions in a Bunch-Davies vacuum during inflation. It would be interesting to see how this exact solution we have obtained above compares to quantum field theory results of fermions in curved backgrounds. Such a comparison may demonstrate that fermions enjoy a preferential status in any background independent formulation of quantum gravity.

\section{Acknowledgments}
   We would like to thank Gabriel Herczeg for fruitful conversations early on in the project. We especially thank João Magueijo for enlightening discussions and for suggesting the inclusion of torsion in the mini-superspace limit of the full fermionic wave function. S.A. thanks the Simons Foundation for funding this project.

\appendix
\appendixpage
\section{Vector and Diffeomorphism Constraints}

Here we apply the (symmetry-reduced) constraints to act on the new state we have derived. Recall the vector and diffeomorphism constraints in the presence of fermionic matter:
\begin{equation}
    \frac{i}{2\kap}{E}^{b}_kF^k_{ab} + {\Pi}^A\cd_a\xi_A = 0, \hspace{0.5cm}\cd_a{E}^{ai}\sigma^{AB}_i + \xi^{(B}{\Pi}^{A)}= 0.
\end{equation}
When we plug our state in these constraints, we find
\begin{equation}
    \qty[\frac{i}{2\kap}\hat{E}^{b}_k{F}^k_{ab} + \widehat{\Pi}^A\cd_a\xi_A]\Psi[A,\xi] = i\hbar\alpha^A\cd_a\xi_A\Omega[A]\psi_K[A]e^{\alpha_A\xi^A},
\end{equation}
\begin{equation}
    \begin{split}
        &\qty[\cd_a\widehat{E}^{ai}\sigma^{AB}_i + \xi^{(B}\widehat{\Pi}^{A)}]\Psi[A,\xi] \\ &= i\hbar\qty[\sigma^{AB}_a\qty(\alpha_E\p_a\xi^E\qty(\frac{\cd_b\xi_C\sigma^{CD}_b\alpha_D}{24\pi GA^2} - \frac{9iV_cA^2}{\Lambda\ell^2_{\rm Pl}}) + \frac{\cd_a\cd_b\xi_C\sigma^{CD}_b\alpha_D}{24\pi GA^2}) + \xi^{(B}\alpha^{A)}]\Psi[A,\xi],
    \end{split}
\end{equation}
where $V_c$ is the volume of the 3D hypersurface. 

\bibliography{quantum_gravity_proj}

\end{document}